\documentclass[prd,twocolumn,showpacs,amsmath,amssymb,nofootinbib]{revtex4-2}
\usepackage{epsf}
\usepackage{graphicx}
\usepackage{epsfig}
\usepackage{xcolor}
\usepackage{subfigure}
\usepackage{pstricks}
\usepackage{pst-node}
\usepackage{rotating}
\usepackage{graphics}
\usepackage{latexsym}
\usepackage{dcolumn}
\usepackage{bm}
\usepackage{times}
\usepackage{indentfirst}
\usepackage{float}
\usepackage{overpic}
\usepackage{amsmath}
\usepackage{appendix}
\usepackage{upgreek}
\usepackage{chngcntr}
\usepackage{amssymb}%
\usepackage{pifont}%
\usepackage[top=0.5in,bottom=0.8in,left=0.9in,right=0.9in]{geometry}
\usepackage{subfigure}
\usepackage{lineno}

\usepackage{multirow}
\usepackage{makecell}

\usepackage{hyperref}
\hypersetup{
	colorlinks=true,
	linkcolor=blue,
	filecolor=blue,
	urlcolor=blue,
	citecolor=blue,
}

\newcommand{\ppp}{\pi^+\pi^- \pi^0}
\newcommand{\pip}{\pi^+}
\newcommand{\pim}{\pi^-}
\newcommand{\piz}{\pi^0}
\newcommand{\kap}{K^+}
\newcommand{\kam}{K^-}
\newcommand{\ks}{K^0_S}
\newcommand{\kl}{K^0_L}

\newcommand{\EE}{e^+e^-}

\newcommand{\GG}{\gamma\gamma}

\newcommand{\pp}{\pi^+\pi^-}
\newcommand{\kk}{K^+K^-}

\newcommand{\gpi}{\gamma\pi^0}

\def\TeV{\ifmmode {\mathrm{\ Te\kern -0.1em V}}\else
                   \textrm{Te\kern -0.1em V}\fi}%
\def\GeV{\ifmmode {\mathrm{\ Ge\kern -0.1em V}}\else
                   \textrm{Ge\kern -0.1em V}\fi}%
\def\MeV{\ifmmode {\mathrm{\ Me\kern -0.1em V}}\else
                   \textrm{Me\kern -0.1em V}\fi}%
\def\keV{\ifmmode {\mathrm{\ ke\kern -0.1em V}}\else
                   \textrm{ke\kern -0.1em V}\fi}%
\def\eV{\ifmmode  {\mathrm{\ e\kern -0.1em V}}\else
                   \textrm{e\kern -0.1em V}\fi}%

\def\TeVc{\ifmmode {\mathrm{\ Te\kern -0.1em V}/c}\else
                   {\textrm{Te\kern -0.1em V}/$c$}\fi}%
\def\GeVc{\ifmmode {\mathrm{\ Ge\kern -0.1em V}/c}\else
                   {\textrm{Ge\kern -0.1em V}/$c$}\fi}%
\def\MeVc{\ifmmode {\mathrm{\ Me\kern -0.1em V}/c}\else
                   {\textrm{Me\kern -0.1em V}/$c$}\fi}%
\def\keVc{\ifmmode {\mathrm{\ ke\kern -0.1em V}/c}\else
                   {\textrm{ke\kern -0.1em V}/$c$}\fi}%
\def\eVc{\ifmmode  {\mathrm{\ e\kern -0.1em V}/c}\else
                   {\textrm{e\kern -0.1em V}/$c$}\fi}%

\def\TeVcc{\ifmmode {\mathrm{\ Te\kern -0.1em V}/c^2}\else
                   {\textrm{Te\kern -0.1em V}/$c^2$}\fi}%
\def\GeVcc{\ifmmode {\mathrm{\ Ge\kern -0.1em V}/c^2}\else
                   {\textrm{Ge\kern -0.1em V}/$c^2$}\fi}%
\def\MeVcc{\ifmmode {\mathrm{\ Me\kern -0.1em V}/c^2}\else
                   {\textrm{Me\kern -0.1em V}/$c^2$}\fi}%
\def\keVcc{\ifmmode {\mathrm{\ ke\kern -0.1em V}/c^2}\else
                   {\textrm{ke\kern -0.1em V}/$c^2$}\fi}%
\def\eVcc{\ifmmode  {\mathrm{\ e\kern -0.1em V}/c^2}\else
                   {\textrm{e\kern -0.1em V}/$c^2$}\fi}%

\let\gevcc=\GeVcc

\def\cm{\ifmmode  {\mathrm{\ cm}}\else
                   \textrm{~cm}\fi}%
\newcommand{\jpsi}{J/\psi}

		\newcommand{\bfg}{\begin{figure}}
			\newcommand{\efg}{\end{figure}}
		\newcommand{\bitm}{\begin{itemize}}
			\newcommand{\eitm}{\end{itemize}}
		\newcommand{\bnum}{\begin{enumerate}}
			\newcommand{\enum}{\end{enumerate}}
		\newcommand{\btbl}{\begin{table}}
			\newcommand{\etbl}{\end{table}}
		\newcommand{\btbu}{\begin{tabular}}
			\newcommand{\etbu}{\end{tabular}}
		\newcommand{\bcl}{\begin{center}}
			\newcommand{\ecl}{\end{center}}
		
		\newcommand{\beq}{\begin{equation}}
			\newcommand{\eeq}{\end{equation}}
		\newcommand{\beqr}{\begin{eqnarray}}
			\newcommand{\eeqr}{\end{eqnarray}}
\begin{document}
	
\title{\boldmath Search for the rare decay $J/\psi \to \gamma D^0+c.c.$ at BESIII}

\author{
M.~Ablikim$^{1}$, M.~N.~Achasov$^{4,c}$, P.~Adlarson$^{76}$, O.~Afedulidis$^{3}$, X.~C.~Ai$^{81}$, R.~Aliberti$^{35}$, A.~Amoroso$^{75A,75C}$, Q.~An$^{72,58,a}$, Y.~Bai$^{57}$, O.~Bakina$^{36}$
% ...
\\
\vspace{0.2cm}
(BESIII Collaboration)\\
\vspace{0.2cm} {\it
$^{1}$ Institute of High Energy Physics, Beijing 100049, People's Republic of China\\
% ...
$^{3}$ Bochum  Ruhr-University, D-44780 Bochum, Germany\\
$^{4}$ Budker Institute of Nuclear Physics SB RAS (BINP), Novosibirsk 630090, Russia\\
% ...
$^{35}$ Johannes Gutenberg University of Mainz, Johann-Joachim-Becher-Weg 45, D-55099 Mainz, Germany\\
$^{36}$ Joint Institute for Nuclear Research, 141980 Dubna, Moscow region, Russia\\
% ...
$^{57}$ Southeast University, Nanjing 211100, People's Republic of China\\
$^{58}$ State Key Laboratory of Particle Detection and Electronics, Beijing 100049, Hefei 230026, People's Republic of China\\
% ...
$^{72}$ University of Science and Technology of China, Hefei 230026, People's Republic of China\\
% ...
$^{75}$ University of Turin and INFN, (A)University of Turin, I-10125, Turin, Italy; (B)University of Eastern Piedmont, I-15121, Alessandria, Italy; (C)INFN, I-10125, Turin, Italy\\
$^{76}$ Uppsala University, Box 516, SE-75120 Uppsala, Sweden\\
% ...
$^{81}$ Zhengzhou University, Zhengzhou 450001, People's Republic of China\\
\vspace{0.2cm}
$^{a}$ Deceased\\
% ...
$^{c}$ Also at the Novosibirsk State University, Novosibirsk, 630090, Russia\\
% ...
}
}

\date{\today}

\renewcommand{\abstractname}{}
\begin{abstract}
Using $(10087\pm44)\times10^6\jpsi$ events collected with the BESIII detector, we search for the rare decay $J/\psi \to \gamma D^0+c.c.$ for the first time. No obvious signal is observed and the upper limit on the branching fraction is determined to be ${\cal B}(\jpsi \to \gamma D^{0}+c.c.)< 9.1 \times 10^{-8}$ at 90\% confidence level.
\end{abstract}

\maketitle

\section{INTRODUCTION} 
\label{sec: intro} 

Search for physics beyond the Standard Model (SM) is one of the major tasks in particle physics. 
Weak decays of the charmonium state $\jpsi$ are an excellent probe of potential New Physics (NP) effects.
In the SM, a $\jpsi$ can weakly decay through Flavor Changing Charged Currents (FCCCs), allowed at tree
level, producing a single charm meson alongside other non-charmed mesons or leptons. These decays have
branching fractions (BFs) in the range of $10^{-9}$ to $10^{-11}$ in the SM~\cite{bran1,bran2,Shen:2008zzb,Dhir:2009rb,Ivanov:2015woa,Wang:2016dkd,Sun:2023uyn}. Many NP models predict an enhancement of these BFs by 2 or 3 orders of magnitude~\cite{topcolour,minsuper,twohiggs}. Therefore, extensive studies of these BFs in the BESIII experiment~\cite{Dh,Dev,Duv,li2024rare} can constrain the parameter space of various models.

\begin{figure}[htbp]
  \centering
  \vspace{-0.5cm}
  \includegraphics[width=\linewidth]{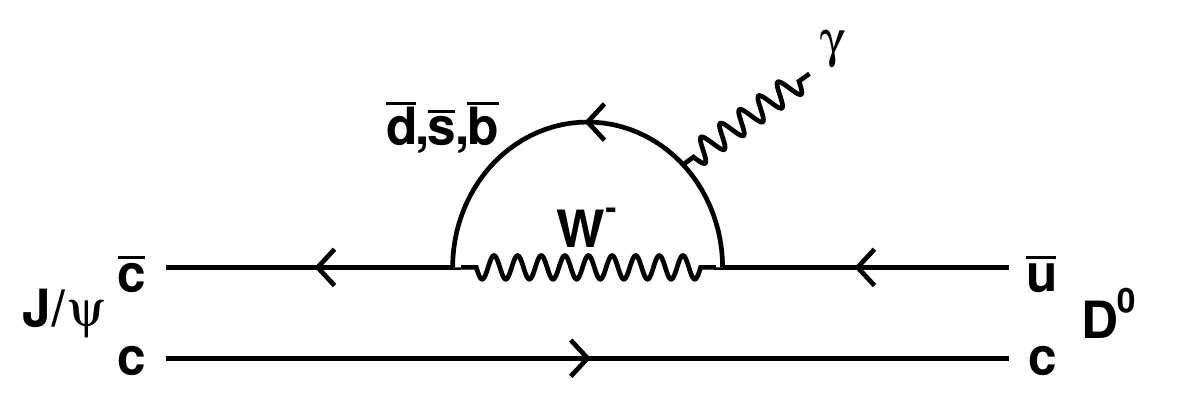}
  \caption{Feynman diagram for $ J/\psi \rightarrow \gamma D^{0} $.}
  \label{fig: feyman}
\end{figure}

In addition, the weak decay of $\jpsi$ mediated by a Flavor Changing Neutral Current (FCNC) is another interesting process. While prohibited at the tree level, it can occur via $c \to u$ transition at the loop level in the SM but is highly suppressed due to the Glashow-Iliopoulos-Maiani (GIM) mechanism~\cite{gim}.
Such decays can also occur via long-distance effects, which are expected to have the same order of magnitude as the production rate of the FCNC process~\cite{long}. 
The FCNC decay $\jpsi \to D^0 l^+ l^-$, within the SM framework, is anticipated to have a BF of the order of $10^{-13}$~\cite{long}. In comparison, the FCNC process $\jpsi \to \gamma D^{0}$ (the charged conjugate channel is always implied throughout the text), as depicted in Fig.~\ref{fig: feyman}, is expected with a 1 or 2 orders of magnitude larger BF, due to the presence of one fewer decay vertex.
The experimental evidence for this FCNC process provides an opportunity to study the non-perturbative QCD effects and their underlying dynamics. Any enhancement of the BF with respect to the SM would be a strong indication of NP~\cite{newo, newth}.

Experimentally, the FCNC processes have been comprehensively studied in the $B$ meson sector via $b \to s$ transitions~\cite{blhcb,bbelle} and in the $D_{(s)}$ meson sector via $c \to u$ transitions~\cite{BESIII:2018hqu,LHCb:2020car,BaBar:2011ouc,BESIII:2021slf,besiii2024search}. So far, they have not been observed in the charm sector. 
The $c \to u \gamma$ transition can be directly probed by studying the radiative decay $\jpsi \to \gamma D^{0}$.
In 2017, the BESIII experiment searched for $\jpsi \to D^0\EE$~\cite{dee} using $(1310.6\pm 7.2) \times 10^6$ $\jpsi$ events resulting in an upper limit on the BF $< 8.5 \times 10^{-8}$ at 90\% confidence level (C.L.). 

In this paper, we search for $\jpsi \to \gamma D^{0}$ based on $(10087 \pm 44) \times 10^6 \jpsi$ events~\cite{jpsieve} collected by the BESIII detector. This is the world's largest $\jpsi$ sample produced at an electron-positron collider. Its clean environment provides an excellent opportunity to search for this rare decay. To study the decay $\jpsi \to \gamma D^{0}$, the $D^{0}$ is reconstructed through its three prominent exclusive hadronic decay modes, $\kam\pip$ (Mode I), $\kam\pip\piz$ (Mode II), and $\kam\pip\pp$ (Mode III), which have large BFs.
 
\section{BESIII DETECTOR AND MONTE CARLO SIMULATION}
\label{sec: detmc}
The BESIII detector~\cite{Ablikim:2009aa} records symmetric $e^+e^-$ collision events
provided by the BEPCII storage ring~\cite{Yu:IPAC2016-TUYA01}
in the center-of-mass energy ranging from 1.84 to 4.95~GeV,
with an achieved peak luminosity of $1.1 \times 10^{33}\;\text{cm}^{-2}\text{s}^{-1}$  at $\sqrt{s} = 3.773\;\text{GeV}$. 
BESIII has collected large data samples in this energy region~\cite{Ablikim:2019hff}. The cylindrical core of the BESIII detector covers 93\% of the full solid angle and consists of a helium-based
 multilayer drift chamber~(MDC), a plastic scintillator time-of-flight
system~(TOF), and a CsI(Tl) electromagnetic calorimeter~(EMC),
which are all enclosed in a superconducting solenoidal magnet
providing a 1.0~T magnetic field. The magnetic field was 0.9~T in 2012, which affects 11\% of the total $J/\psi$ data.
The solenoid is supported by an
octagonal flux-return yoke with resistive plate counter muon
identification modules interleaved with steel. 
The momentum resolution of charged-particle at $1~{\rm GeV}/c$ is
$0.5\%$, and the  ${\rm d}E/{\rm d}x$ resolution is $6\%$ for electrons from Bhabha scattering. 
The EMC measures photon energies with a
resolution of $2.5\%$ ($5\%$) at $1$~GeV in the barrel (end cap)
region. The time resolution in the TOF barrel region is 68~ps, while
that in the end cap region is 110~ps. The end cap TOF
system was upgraded in 2015 using multigap resistive plate chamber
technology, providing a time resolution of 60~ps, which benefits 87\% of the data used in this analysis~\cite{tof1,tof2,tof3}.

Simulated data samples produced with a {\sc geant4}-based~\cite{geant4} Monte Carlo (MC) package, 
which includes the geometric description of the BESIII detector and the
detector response, are used to determine the detection efficiencies
and to estimate backgrounds. In the simulation, the beam
energy spread and initial state radiation in the $e^+e^-$
annihilations are modeled with the generator {\sc kkmc}~\cite{ref:kkmc}. 
An inclusive MC sample of $10^{10}$ $\jpsi$ events, including the production of the $J/\psi$ resonance, is generated to study the backgrounds. 
All particle decays are modeled either with {\sc evtgen}~\cite{ref:evtgenl, ref:evtgenp} using BFs  taken from the Particle Data Group (PDG)~\cite{pdg}, when available,
or otherwise estimated with the {\sc lundcharm}~\cite{ref:lundcharm1, ref:lundcharm2} package.
Final state radiation
from charged final state particles is incorporated using the {\sc
photos} package~\cite{photos}. To estimate the detection efficiency, the signal events of $\jpsi \to \gamma D^{0}$ are generated by the JPE model which is constructed for a vector meson decays into a photon plus a pseudoscalar meson, with the phase space model for $D^{0} \to\kam\pip$ and the model from the amplitude analyses for $D^{0}\to\kam\pip\piz$ and $D^{0}\to\kam\pip\pp$~\cite{Modelk3pi}.

\section{DATA ANALYSIS}
\label{dataana}

\subsection{Common selection}
\label{sub: comsel}
Charged tracks reconstructed in the MDC are required to be within a polar angle ($\theta$) range of $|\rm{cos\theta}|<0.93$, where $\theta$ is defined with respect to the $z$-axis,
which is the symmetry axis of the MDC. The distance of closest approach to the interaction point (IP) must be less than 10\,cm along the $z$-axis, and less than 1\,cm in the transverse plane. Particle identification~(PID) is carried out for charged tracks by combining the specific ionization energy loss measured by the MDC~(d$E$/d$x$) and the flight time in the TOF to form the likelihoods $\mathcal{L}(h)$ for the $h=K,~\pi$ hypotheses. Charged kaons and pions are identified by requiring $\mathcal{L}(K)>\mathcal{L}(\pi)$ and $\mathcal{L}(\pi)>\mathcal{L}(K)$, respectively.

Photon candidates are identified by using showers in the EMC. The deposited energy of each shower must be more than 25~MeV in the barrel region ($|\cos \theta|< 0.80$) and more than 50~MeV in the end cap region ($0.86 <|\cos \theta|< 0.92$). To exclude showers originating from charged tracks, the opening angle subtended by the EMC shower and the position of the closest charged track at the EMC must be greater than 20 degrees as measured from the IP. To suppress electronic noise and showers unrelated to the event, the difference between the EMC time and the event start time is required to be within [0, 700]\,ns.
The $\piz$ candidates are reconstructed with pairs of selected photons. The invariant mass of the photon pair is required to be within [0.115, 0.150]\,$\GeVcc$. To improve the kinematic resolution, a kinematic fit~\cite{kine} is carried out, constraining the invariant mass of the photon pair to be the $\piz$ nominal mass~\cite{pdg}, and the corresponding $\chi^2$ of the kinematic fit is required to be less than 20.


The candidate events are selected by requiring a radiative photon from $\jpsi$ decay, the charged track of a kaon, and one or three charged pions (depending on the $D^0$ decay modes). The radiative photon with the maximum energy is selected and the charged tracks are required to have zero net charge. All three modes require to have no additional charged track, and Mode II requires one $\piz$ candidate.
To improve the resolution and reduce the backgrounds, a four-constraint (4C) kinematic fit is performed, imposing energy-momentum conservation under the hypothesis of $\jpsi \to \kam\pip\gamma$ for Mode I or $\jpsi \to \kam\pip\pp\gamma$ for Mode III. Alternatively, a five-constraint (5C) kinematic fit is carried out under the hypothesis of $\jpsi \to \kam\pip\gamma\gamma\gamma$ for Mode II, with an additional constraint on the $\gamma\gamma$ invariant mass to match the $\piz$ nominal mass. For the fit, $\chi^2_{\rm 4C/5C}<30$ is required.
For events with more than one $\piz$ candidate in Mode II, the one with the minimum $\chi^2_{\rm 5C}$ value is retained for further analysis. The momenta of the charged tracks and the showers after the kinematic fit are used for further analysis throughout the text if not mentioned explicitly.

\subsection{Further selection}
\label{sub: fursel}
For Mode I, the dominant backgrounds are  $\jpsi \to \pp\gamma$, $\jpsi \to \kk\gamma$, and $\jpsi \to \ppp$, due to the misidentification between a high momentum kaon and a pion, or due to missing a soft photon.
To remove these backgrounds, 4C kinematic fits under the hypotheses of $\jpsi \to \pp\gamma$, $\jpsi \to \kk\gamma$ and $\jpsi \to \pp\GG$ (named as Alternative kinematic fit1/fit2/fit3, similarly named for Mode II and III) are carried out demanding $\chi^2_{\pi^+\pi^-\gamma/K^+K^-\gamma/\pi^+\pi^-\gamma\gamma}>85/50/35$. These requirements have been optimized by maximizing the quantity of Punzi FOM $\epsilon/(1.5+\sqrt{B})$~\cite{punzi} (analogously employed in the similar event selection), where $\epsilon$ is the signal efficiency and $B$ is the background yield. After these selections, the momentum requirements, $p_{K/\pi}<1.25/1.35~\GeVc$, are applied to suppress the backgrounds from $K/\pi$ misidentification.
Other backgrounds from radiative Bhabha and di-muon events occur due to misidentifying leptons as hadrons and their large cross sections. To suppress the radiative Bhabha background, the requirement of $E/p<0.8$ is applied for any charged pion candidate, where $E$ and $p$ are the corresponding energy deposited in the EMC and the momentum measured in the MDC, respectively. 
The requirement of $\chi^2_{\pi^+\pi^-\gamma}>85$ suppresses the radiative dimuon background.

For Mode II, the dominant background is $\jpsi\to\kam\pip\piz\piz$ with the intermediate process $\ks \to \piz\piz$ with a missing soft photon. A mass window $0.32<M^{\rm Recoil}_{\kam\pip}<0.66~ \GeVcc$ is applied to veto this background, where $M^{\rm Recoil}_{\kam\pip}$ is the recoil mass of the $\kam\pip$ system calculated with the $\kam\pip$ momenta before the kinematic fit. 
There are also backgrounds from  $\jpsi \to \ppp\gamma$ (including $\omega \to \gpi$),  $\jpsi \to \kk\piz\gamma$, and $\jpsi \to \ppp\piz(\gamma\gamma)$ due to the misidentification between a high momentum kaon and a pion, or due to missing a soft photon. 
To suppress these backgrounds, 4C kinematic fits under the hypotheses of $\jpsi \to \pp\gamma\gamma\gamma$, $\jpsi \to \kk\GG\gamma$ and $\jpsi \to \pp\GG\GG$ are performed and $\chi^2_{\pi^+\pi^-\piz\gamma/K^+K^-\piz\gamma/\pip\pim\piz\gamma\gamma}>40/40/95$ are required, individually. 
A mass window of the $\gpi$ invariant mass, $0.68<M_{\gpi}<0.92\ \GeVcc$, is applied to veto the background from $\jpsi \to \omega\pp \to \ppp \gamma$. 
Another potential background is $\jpsi \to \kk\gamma$ due to the charged kaon decay $\kap \to \pip\piz$.  A mass window is applied to the $\pip\piz$ invariant mass with $0.45<M_{\pip\piz}<0.53~\gevcc$ to veto this background. 
The background with $\jpsi\to\kam\pip\piz\kl$ also presents in the case of a $\kl$ shower in the EMC misidentified as the radiative photon candidate. 
To veto this background, a Multivariate Data Analysis (MVA) in the ROOT TMVA package~\cite{tmva} is performed on the shower shape in the EMC based on a  Gradient Boosted Decision Trees (BDTG) algorithm. 
The input parameters in the MVA include $N_{\rm hit}$ (the number of hit crystals in the EMC), $E_{\rm seed}/E_{3\times3}$ (ratio  of energy deposited in the center crystal of the shower to that in the 3$\times$3 crystals), $E_{3\times3}/E_{5\times5}$ (ratio of energy deposited in the 3$\times$3 crystals to that in the 5$\times$5 crystals) and $A_{42}$ moment as defined in Ref.~\cite{pinet} with relatively small correlation. In order to train and test the MVA, the photon candidates from the exclusive signal MC sample and the $\kl$ candidates from the exclusive MC sample of $\jpsi \to \kam\pip\piz\kl$  are used. By applying the MVA method on the surviving events of inclusive and signal MC samples, the $\kl$ background can be suppressed by 90\% with the requirement of the BDTG output value.

For Mode III, the dominant background is $\jpsi \to \kam\pip\pp\gamma(\piz)$ with a $\pp$ pair from $\ks$ decay (missing a soft photon).
Therefore a mass window $0.46<M_{\pp}<0.55\ \GeVcc$ is applied to veto this background, where $M_{\pp}$ is the invariant mass of any $\pp$ pair (two pairs per event). 
Another main background contribution is $\jpsi \to \pp\pp\gamma$ due to the misidentification between a high momentum kaon and a pion. 
A 4C kinematic fit under the hypothesis of $\jpsi \to \pp\pp\gamma$ is carried out and $\chi^2_{\pi^+\pi^-\pip\pim\gamma}>430$ is required. 
There are also backgrounds due to charged kaon decay, such as $\jpsi\to\kk\gamma$ with $\kap\to\pip\pp$ and $\jpsi\to\kk\pp$ with $\kap\to\pip\piz$ and a missing soft photon. The corresponding mass windows for the $\pip\pp$ invariant mass $0.47<M_{\pip\pp}<0.52\ \GeVcc$ and the recoil mass of the $\kam\pp$ combination with the least deviation from the nominal $K$ mass $0.42<M^{\rm Recoil}_{\kam\pp}<0.52\ \GeVcc$ are applied to veto these backgrounds.
 Another potential background is $\jpsi \to \kk\piz$ due to the Dalitz decay $\piz\to\EE\gamma$. To reject this background, a mass window of the $\EE\gamma$ invariant mass (replacing the mass of the pion with that of an electron), $0.09<M_{\EE\gamma}<0.19\ \GeVcc$, is applied.

\subsection{Signal extraction and branching fraction calculation}
\label{sub: sigext}
\begin{figure*}[htbp]
  \center
  \begin{overpic}[width=0.95\textwidth]{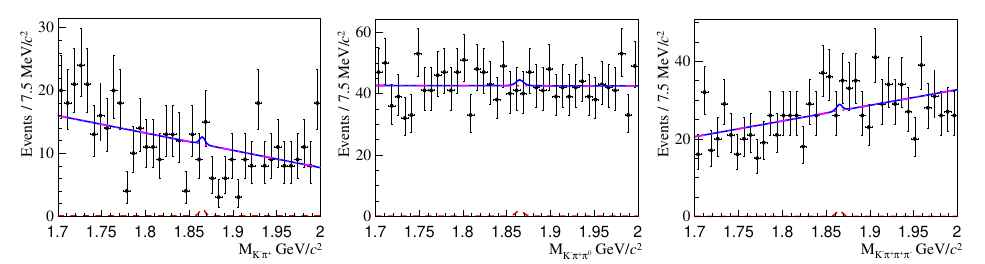}
  \put(9, 7){$(a)$}
  \put(42,7){$(b)$}
  \put(75,7){$(c)$}
  \end{overpic}
  \vskip -0.3cm
  \caption{The unbinned simultaneous fit on the invariant mass distributions of Modes $(a)$ I, $(b)$ II and $(c)$ III, where the dots with error bar are data, the blue lines are the fit result, and the red dotted and purple curves are the fitted signal and background, respectively.}
  \label{fig: signalsearchall}
\end{figure*}

After the selection criteria described in Sec~\ref{sub: fursel} have been applied, the invariant mass distributions of the $\kam\pip$, $\kam\pip\piz$ and $\kam\pip\pp$ for the three $D^0$ decay modes are shown in Fig.~\ref{fig: signalsearchall}.
The BF of $ J/\psi \rightarrow \gamma D^{0} $ is calculated by 
\begin{eqnarray}
  {\cal B}(J/\psi \rightarrow \gamma D^{0} + c.c.) = \frac{N^{\text{sig}}_i}{N_{\jpsi} \times {\cal B}_i\times \epsilon_{i}},
  \label{eq: branch}
\end{eqnarray}  
where $N^{\text{sig}}_{i}$ is the signal yield observed in the $i$-th decay mode, $N_{\jpsi}$ is the total number of $\jpsi$ events, ${\cal B}_i$ is the corresponding BF of the $D^0$ decay and its subsequent daughter particles' decays, and $\epsilon_i$ is the reconstruction efficiency obtained with the signal MC samples, which is $(22.96 \pm 0.11)\%$, $(14.83 \pm 0.09)\%$, and $(17.55 \pm 0.10)\%$ for Modes I, II and III, respectively.

To extract the BF of $\jpsi\to\gamma D^0$, an unbinned simultaneous maximum-likelihood fit is carried out to the selected samples of the three $D^0$ decay modes by sharing the same decay BF of  $\jpsi\to\gamma D^0$. In the fit, the signal is described with the simulated shape derived from the signal MC sample, and the background is described with a line, with all parameters allowed to float.
The simultaneous fit results are shown in Fig.~\ref{fig: signalsearchall}. No obvious signal is observed, therefore an upper limit on the BF of  $\jpsi\to\gamma D^0$ at 90\% C.L. is set in Sec~\ref{sub: upp}.

\subsection{Systematic uncertainties}
\label{sub: sys} 
Several systematic uncertainties are considered in the BF measurement.
The uncertainties are separated into two categories according to the approaches to evaluate their effects on the BF measurement, 
1) effects that are evaluated with the different scenarios (under different selection criteria) and extracted by their resultant differences with respect to the nominal value, named category A thereafter; 
2) effects that are evaluated with the various methods, and propagated to the final results directly, named category B thereafter.
The sources of systematic uncertainties and the assigned values are summarized in Table~\ref{tab: sumunsec}.


The category A uncertainties include those associated with the $K/\pi$ momentum requirements, the mass window veto requirement as well as the shapes of signal and background in the fit. 
The uncertainties associated with the $K/\pi$ momentum requirements 
are estimated by changing the boundary by $\pm0.005$ or $\pm 0.010~\GeVc$ based on the nominal values.
The different mass windows are applied to veto the backgrounds associated with narrow intermediate states $\ks$, $K$, $\omega$ and $\piz$. 
The corresponding uncertainties are estimated by varying the mass window boundaries by $\pm1$ or $\pm 2$ times the corresponding mass resolution.
The uncertainties associated with the line shapes of signal and background in the fit are estimated with the alternative line shapes, $i.e.$, the signal shape is parameterized with a double Gaussian function with the parameters extracted from a fit to the signal MC sample, while the background line shape is described with a second-order polynomial function.
All above alternative scenarios are carried out, and the differences with respect to the nominal value are taken as the corresponding systematic uncertainties. 


The category B uncertainties are those associated with the tracking and PID for the charged tracks, the photon detection and $\piz$ reconstruction, the $E/p$ ratio, the MVA cut for the radiative photon, the number of extra track requirement, the kinematic fit, the quoted BFs of intermediate states, and the total number of $\jpsi$ events. 

To estimate the corresponding uncertainties, the efficiencies of tracking and PID for the pions and koans are studied with the control samples of $\jpsi\to\ppp$ and $\jpsi\to\ks\kam\pip\to\kam\pip\pp$, respectively. The efficiencies of photon detection~\cite{gamun}, $\piz$ reconstruction~\cite{gamun} and $E/p$ ratio are studied with control sample of  $\jpsi\to\ppp$ individually.
All above efficiencies are studied both for the data and for the corresponding MC samples in bins of $p$ and cos$\theta$. The average relative differences of the efficiencies between data and MC simulation, which are obtained by weighting according to the distribution of signal MC samples, are taken as the systematic uncertainties.



The efficiencies of the number of extra tracks requirement and the MVA cut for the radiative photon are also studied using the control sample of $\jpsi\to\ppp$. The relative differences of the efficiencies between data and MC simulation are taken as the uncertainty. 
The uncertainty associated with the kinematic fit is mainly due to the inconsistency of the kinematic variables of charged tracks. Therefore the kinematic variables of charged tracks of the signal MC sample are smeared to minimize the difference between data and MC simulation, and the resulting change in the detection efficiency is taken as the uncertainty.
The systematic uncertainties due to the kinematic fit are estimated for all the signal and background hypotheses individually. The details of smearing parameters of charged tracks can be found in Ref.~\cite{kincor}.
The uncertainties of the decay BFs of intermediate states are taken from the PDG~\cite{pdg}. The uncertainty of the total number of $\jpsi$ events is taken from Ref.~\cite{jpsieve}.

\begin{table}[htbp]
	\centering
	\caption{Sources of systematic uncertainties and the corresponding uncertainties to the BF (in \%), where ``-'' denotes not applied, and ``*'' indicates the category A uncertainties.}
   \resizebox{\linewidth}{!}{
	\begin{ruledtabular}
	\begin{tabular}{c|cccc}
		Source/Mode & I & II & III \\ \hline

        Momentum      & * & - & - \\
		Veto $K_S^0$  & - & * & * \\
		Veto $K$ decay  & - & * & * \\	 
		Veto $\omega$ & - & * & - \\
		Veto $\pi^0$  & - & - & * \\ 
		Signal lineshape  & * & * & * \\
		Background lineshape  & * & * & * \\ \hline

		Tracking & 0.6 & 1.1 & 2.4 \\
		PID & 2.0 & 1.4 & 1.8 \\
		Photon requirement & 1.0 & 1.0 & 1.0 \\
		$\pi^0$ selection & - & 2.2 & - \\
		$N_{\rm extra}=0$ & 0.1 & 0.1 & 0.1 \\
		Kinematic fit & 0.8 & 0.5 & 2.0 \\ 
		Alternative kinematic fit1 & 0.2 & 0.4 & 0.8 \\
		Alternative kinematic fit2 & 0.0 & 0.0 & - \\
		Alternative kinematic fit3 & 0.1 & 0.1 & - \\
        $E/p$ ratio & 0.9 & - & - \\
		MVA cut   & - & 0.2 & - \\  
        Quoted BFs & 0.8 & 4.2 & 1.7 \\
        $N_{\jpsi}$ & 0.4 & 0.4 & 0.4 \\ \hline  
	    Total & 2.7 & 5.2 & 4.1 \\ 
	\end{tabular}
	\end{ruledtabular}
	}
	\label{tab: sumunsec}
\end{table}

\subsection{Upper limit on branching fraction}
\label{sub: upp}
\begin{figure}[htbp]
	\center
	\includegraphics[width=0.45\textwidth]{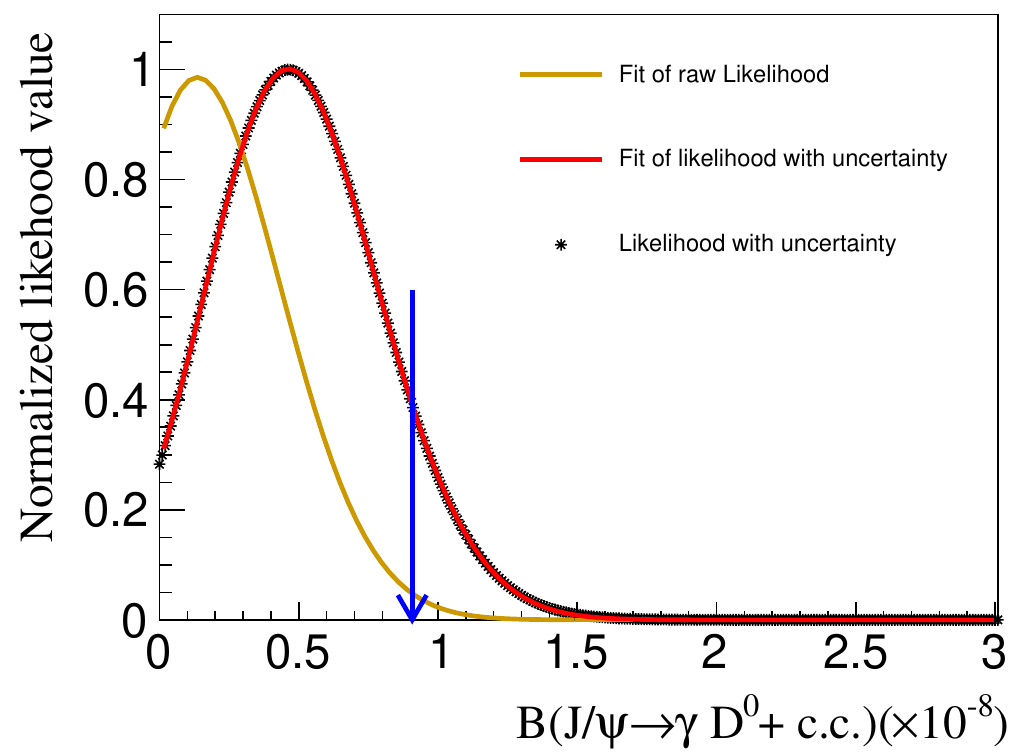}
	\caption{The likelihood profile for ${\cal B}(\jpsi \to \gamma D^{0} + c.c.)$ combining the three decay modes of the three decay modes to extract the upper limit on the BF. The black dots are the likelihood distribution with systematic uncertainty, the blue arrow is the 90\% integral value of the red line, and the red and orange lines are the curves of the fit results for the likelihood with and without category A and B uncertainties, respectively.}
	\label{fig: upperlimitall}
\end{figure}

As the data is compatible with a background-only hypothesis, a Bayesian method~\cite{ZHU2007322} is utilized to obtain an upper limit at 90\% C.L. on the BF of $\jpsi\to\gamma D^0$.
The likelihood values are obtained by performing simultaneous fits on the distributions of the invariant masses of $\kam\pip$, $\kam\pip\piz$, and $\kam\pip\pp$ with different BF hypotheses, as shown in Fig.~\ref{fig: upperlimitall}.

To incorporate the effects of category A uncertainties, the alternative simultaneous fits for the three $D^0$ decay modes are carried out with all different scenarios, as discussed in Sec.~\ref{sub: sys}. The fit with the largest upper limit, ensuring the most conservative estimate, is selected for analysis.
To further incorporate the category B uncertainties, the simultaneous fit for the three $D^0$ decay modes is carried out, where the BF is the fitting parameter and the signal efficiencies of the three decay modes are treated as nuisance parameters. The distributions of the signal efficiencies for the three $D^0$ decay modes are assumed to follow a Gaussian distribution with a mean of the corresponding signal MC efficiency, and a width of the corresponding absolute systematic uncertainties.
The profiled likelihood value incorporating the category B uncertainties  is shown in Fig.~\ref{fig: upperlimitall}.


The upper limit at 90\% C.L. on the BF of $\jpsi\to\gamma D^0$ is ${\cal B}(\jpsi \to \gamma D^{0})< 9.1 \times 10^{-8}$ with the systematic uncertainties ($6.1 \times 10^{-8}$ without the systematic uncertainties), obtained by integrating the profile likelihood value from 0 to a definitive value which corresponds to 90\% of total area. 



\section{SUMMARY}
\label{sum}
In summary, we have presented the first search for the rare decay $\jpsi \to \gamma D^0$ using $(10087\pm44)\times10^6\ \jpsi$ events collected with the BESIII detector. Our results are consistent with a background-only hypothesis, and we establish an upper limit on the branching fraction of ${\cal B}(\jpsi \to \gamma D^{0} + c.c.)< 9.1 \times 10^{-8}$ at 90\% C.L., representing the most stringent limit to date. Although our measurement do not reach the precision predicted by the SM, it provides a valuable reference for studying different NP models and restrict the phase space of parameters. 
The sensitivity of searching for this decay could be improved with a larger $\jpsi$ sample, potentially achievable through future experiments~\cite{STCF}.

\begin{acknowledgments}
The BESIII Collaboration thanks the staff of BEPCII and the IHEP computing center and the supercomputing center of the University of Science and Technology of China (USTC) for their strong support. 
This work is supported in part by National Key R\&D Program of China under Contracts Nos. 2020YFA0406400, 2020YFA0406300, 2023YFA1606000, 2023YFA1609400; National Natural Science Foundation of China (NSFC) under Contracts Nos. 11625523, 11635010, 11735014, 11935015, 11935016, 11935018, 12025502, 12035009, 12035013, 12061131003, 12105276, 12122509, 12192260, 12192261, 12192262, 12192263, 12192264, 12192265, 12221005, 12225509, 12235017, 12361141819; the Chinese Academy of Sciences (CAS) Large-Scale Scientific Facility Program; the CAS Center for Excellence in Particle Physics (CCEPP); Joint Large-Scale Scientific Facility Funds of the NSFC and CAS under Contract No. U1832207, U2032111, U1732263, U1832103; CAS Youth Team Program under Contract No. YSBR-101; 100 Talents Program of CAS; The Institute of Nuclear and Particle Physics (INPAC) and Shanghai Key Laboratory for Particle Physics and Cosmology; German Research Foundation DFG under Contracts Nos. 455635585, FOR5327, GRK 2149; Istituto Nazionale di Fisica Nucleare, Italy; Ministry of Development of Turkey under Contract No. DPT2006K-120470; National Research Foundation of Korea under Contract No. NRF-2022R1A2C1092335; National Science and Technology fund of Mongolia; National Science Research and Innovation Fund (NSRF) via the Program Management Unit for Human Resources \& Institutional Development, Research and Innovation of Thailand under Contract No. B16F640076; Polish National Science Centre under Contract No. 2019/35/O/ST2/02907; The Swedish Research Council; U. S. Department of Energy under Contract No. DE-FG02-05ER41374.
\end{acknowledgments}
	
\bibliography{main}
	
\end{document}